\documentclass[sigconf]{acmart}
\usepackage[linesnumbered,ruled,vlined]{algorithm2e}
\AtBeginDocument{%
  \providecommand\BibTeX{{%
    \normalfont B\kern-0.5em{\scshape i\kern-0.25em b}\kern-0.8em\TeX}}}

\acmConference[KDD Workshop '19]{2019 KDD Workshop on Applied Data Science for Healthcare: Bridging the Gap between Data and Knowledge}{August 08, 2019}{Anchorage, AK}
\acmBooktitle{KDD Workshop 2019: Applied Data Science for Healthcare,
 August 08, 2019, Anchorage, AK}
\begin{document}

\title{A Systematic Approach to Detect Hierarchical Healthcare Cost Drivers and Interpretable Change Patterns}


\author{Ta-Hsin Li}
\email{thl@us.ibm.com}
\author{Huijing Jiang}
\email{huijiang@us.ibm.com}
\affiliation{%
\institution{IBM T.J. Watson Research Center}
% \streetaddress{1101 Kitchawan Rd}
  %\city{Yorktown Heights}
  %\state{NY}
  %\postcode{10598}
}

\author{Kevin Tran}
\email{kntran@us.ibm.com}
\author{Gigi Yuen-Reed}
%\author{Thomas Halvorson}
\email{gigi.yuen@us.ibm.com}
%\email{thalvors@us.ibm.com}
\affiliation{%
  \institution{IBM Watson Health}
%  \streetaddress{1 Th{\o}rv{\"a}ld Circle}
%  \country{Iceland}
}
%\email{larst@affiliation.org}
\author{Bob Kelley}
\email{bkelley@us.ibm.com }
\author{Thomas Halvorson}
\email{thalvors@us.ibm.com}
\affiliation{%
	\institution{IBM Watson Health}
	%  \streetaddress{1 Th{\o}rv{\"a}ld Circle}
	%  \country{Iceland}
}

%\affiliation{%
%  \institution{Inria Paris-Rocquencourt}

%\affiliation{%
 %\institution{Rajiv Gandhi University}

%\affiliation{%
%  \institution{Tsinghua University}
%  \streetaddress{30 Shuangqing Rd}

%\affiliation{%
%  \institution{Palmer Research Laboratories}
%  \streetaddress{8600 Datapoint Drive}

%\affiliation{\institution{The Th{\o}rv{\"a}ld Group}}
%\email{jsmith@affiliation.org}

%\affiliation{\institution{The Kumquat Consortium}}
%\email{jpkumquat@consortium.net}

\renewcommand{\shortauthors}{Li and Jiang, et al.}
\renewcommand{\shorttitle}{Detect Hierarchical Healthcare Cost Drivers and Interpretable Change Patterns}
\begin{abstract}
There is strong interest among payers to identify emerging healthcare cost drivers to support early intervention. However, many challenges arise in analyzing large, high dimensional, and noisy healthcare data. In this paper, we propose a systematic approach that utilizes hierarchical and multi-resolution search strategies using enhanced statistical process control (SPC) algorithms to surface high impact cost drivers. Our approach aims to provide interpretable, detailed, and actionable insights of detected change patterns attributing to multiple demographic and clinical factors. We also proposed an algorithm to identify comparable treatment offsets at the population level and quantify the cost impact on their utilization changes.
\end{abstract}




\maketitle

\section{Introduction}
There is strong interest to better understand and manage drivers of healthcare cost. Payers, including public agencies, private health plans, and self-insured employers, are particularly interested in identifying emerging cost drivers to support early intervention. Traditional approaches to identify cost drivers within large payer claims databases can be labor-intensive. It may require manually drilling into the data, and often times, drivers exhibit insidious trends, are masked within summary reports, or are too general to deem actionable.

Data scientists tasked with cost driver detection can be overwhelmed by the sheer volume of data. Their analysis may be driven by personal experience which could bias their approach to perform a systematic and comprehensive detection. Furthermore, several factors make analyzing healthcare claims data a challenge. First, claims data is high dimensional and sparse with noisy signal strength. Second, data collection and standardization can be inconsistent across data sources, due to variation in clinical coding practices, healthcare delivery and payment models, and speed of claims processing. Third, healthcare cost drivers tend to be inter-related, making it difficult to isolate the underlying root cause. Finally, healthcare cost changes could be influenced by seasonal factors, clinical guidelines, or new technologies introduced to the market. These considerations make it challenging for data scientists and payers to determine where to efficiently focus their efforts on.

In this paper, we introduce a statistical process control (SPC)-based detection algorithm to identify emerging healthcare cost drivers. Applications to SPC have been used in many healthcare domains\cite{Thor07, Deborah15, Ray17}. However, existing studies have been limited to a specific outcome within a target population (e.g., inpatient readmission rates for an individual hospital).

Instead of looking at a particular condition or treatment, our approach performs an exhaustive search. We utilize hierarchical and multi-resolution schemes to characterize the temporal change patterns and attribute the changes to multiple explanatory factors. Compared to other cost attribution studies\cite{Katz15}, we leverage clinical episode-based groupings and multi-factor drill downs for better interpretability. We also propose an algorithm to monitor ``utilization offsets'' of comparable treatments and estimate the cost impact attributed to the offset effect. Utilization offsets or treatment switch analyses have been studied extensively within real-world evidence applications\cite{Feldman19}. Our approach extends this concept to evaluate the cost impact of offsets for drug treatments, disease severity, as well as care setting.

%The remainder of the paper is organized as follows. Section \ref{sec:search} introduces our hierarchical and multi-resolution search strategies.  Section \ref{sec:learn} discribes the change detection algorithm and pattern characterization approach. Section \ref{sec:MOD} presents the proposed algorithm for offsets identification and impact qualification. Concluding remarks are given in Section \ref{sec:conclusion}.

%\section{Data}\label{sec:data}

\section{Search Strategies}\label{sec:search}
We apply hierarchical and multi-resolution search strategies leveraging domain knowledge-based factors to search for impactful cost drivers. The claims data is aggregated into multiple key performance indicator (KPI) time series based on (1) different hierarchical drill paths or ``viewpoints'' and (2) time horizon settings.

We use a 1 million random sample of enrollees from the IBM Watson Health MarketScan Commercial Database \cite{marketscan} from 2012-2016. The database includes a longitudinal perspective of enrollment, demographic, medical and pharmacy claims data from large employers and health plans. The claims data provides demographic and clinical information (e.g., diagnoses, procedures, drug codes, care setting, etc.) and cost information associated with the healthcare services. We also utilize medical groupers that combines medical and pharmacy claims into unique episodes of care which allows us to tie drug information to specific conditions \cite{MEG}. Finally, we use the Micromedex RED BOOK database to establish a knowledge base of comparable drug treatments for the offsets identification algorithm\cite{redbook}.

\begin{figure}[h]
	\centering	
	\includegraphics[width=\linewidth]{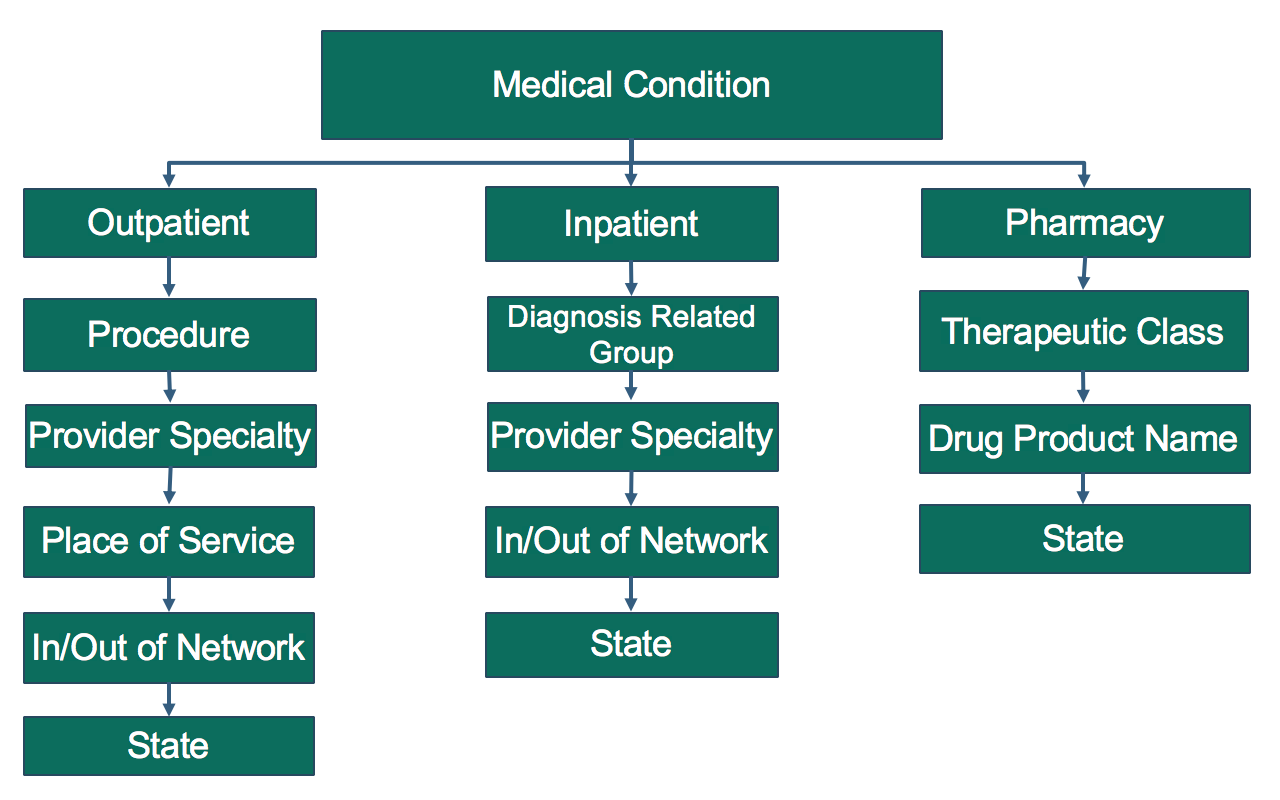}	
	\caption{Example of Viewpoint Hierarchy}	
	\label{fig:hierarchy}	
\end{figure}

The claims data is grouped into clinical episodes of care by assigning an event label (e.g., treatment episodes, admissions) to each claim record. The records are then aggregated according to viewpoint definitions. Figure \ref{fig:hierarchy} shows an example of different viewpoints based on demographic and clinical attributes. For example, a viewpoint pertaining to a specific condition, pharmacy claim type, therapeutic class, and drug product name hierarchy will provide insights on a specific drug for treating a specific condition. This can also be extended to inpatient and outpatient claims and their related attributes (e.g., procedure, provider specialty, place of service, geography, etc.).

\begin{figure}[h]	
	\centering	
	\includegraphics[width=\linewidth]{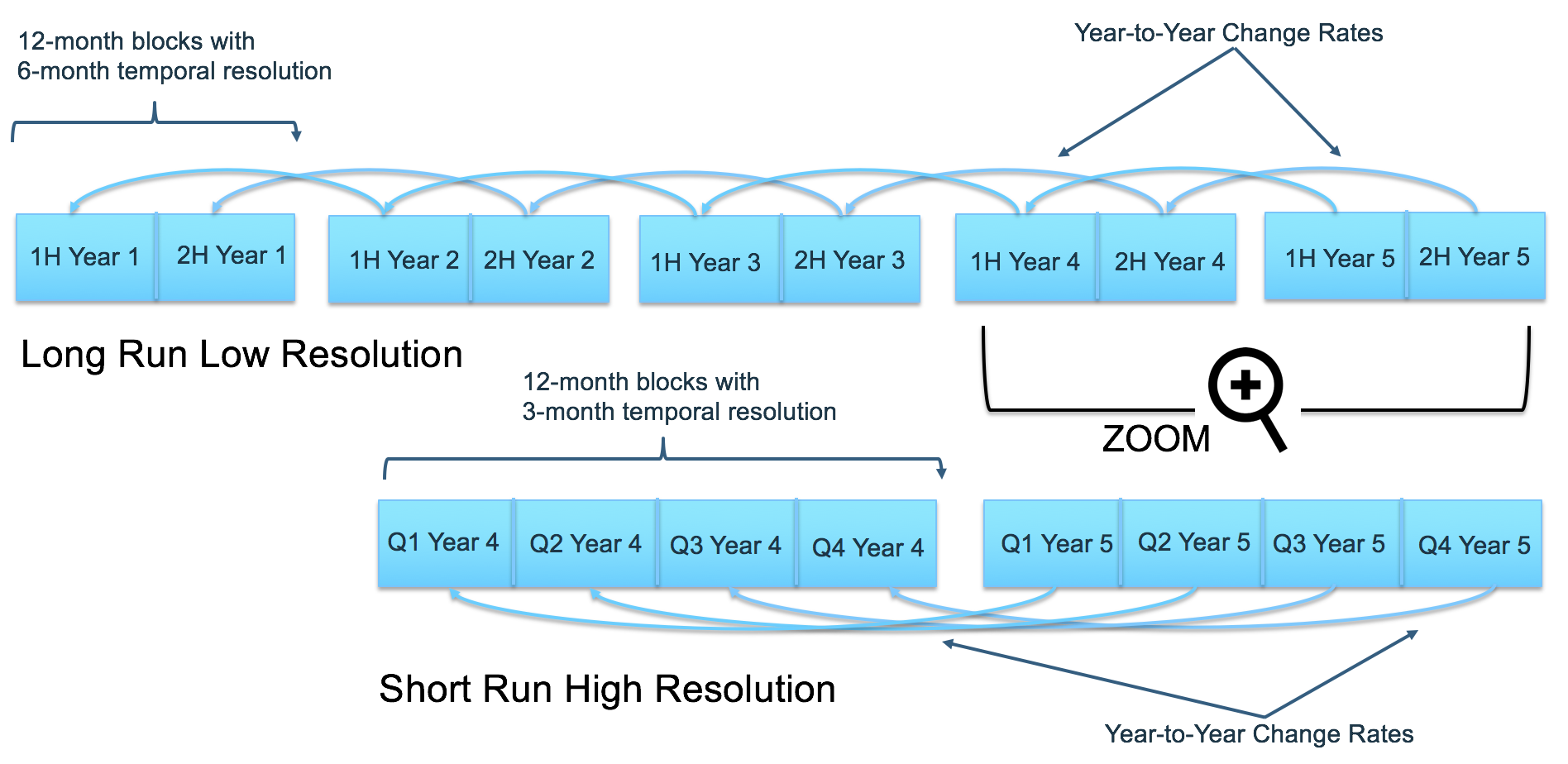}	
	\caption{Example of Multi-Resolution Analysis}	
	\label{fig:resolution}	
\end{figure}

Data are also aggregated temporally using multiple analysis windows with different durations and resolutions as shown in Figure \ref{fig:resolution}. Changes are detected based on year-to-year comparisons of the KPIs while controlling for lags in payment processing over time.

\begin{figure}[h]
	\centering
	\includegraphics[width=\linewidth]{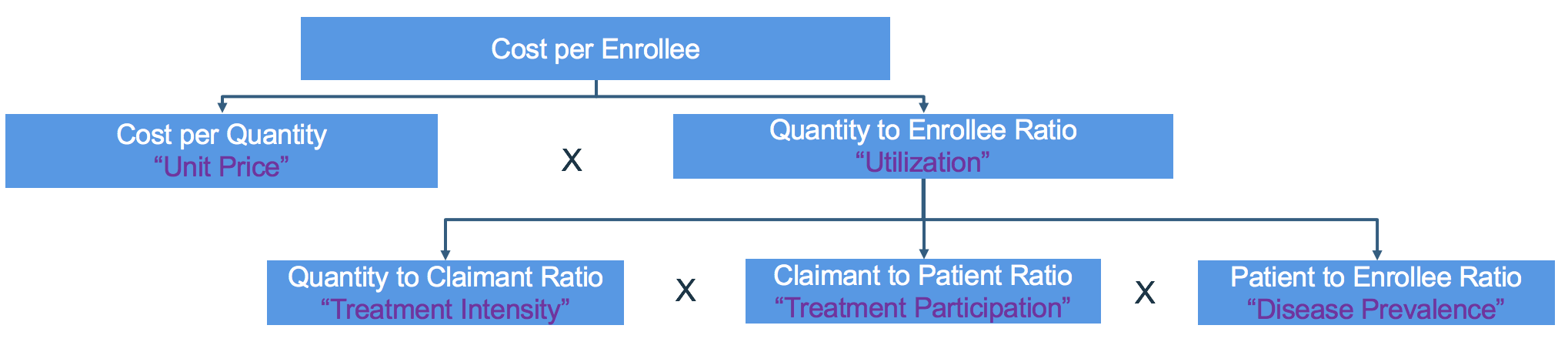}	
	\caption{Example of KPI Decomposition}	
	\label{fig:KPI}	
\end{figure}

Various KPIs are computed for the time series. For each viewpoint and time period combination, we calculate the total cost, number of episodes, number of enrollees, number of patients with a specific condition, number of claimants on a specific treatment, and quantity of services. This provides information to calculate the ratios for each KPI and standard errors shown in Figure \ref{fig:KPI}.

\section{Change Pattern Learning}\label{sec:learn}
%For each viewpoint, we first detect the presence of a significant change in the KPI time series for each driver by different analysis types. We then apply classification methods to profile the detected change patterns.
\subsection{Detect Change Patterns}\label{sec:detect}
To detect changes early, reliably, and in a clinically meaningful manner, we developed an enhanced SPC-based analytics algorithm incorporating domain hierarchical knowledge. It contains three primary functionalities: a threshold learning module to establish a detection threshold through historical data modeling, an online detection mechanism (e.g., auto-reset or non-restarting CUSUM \cite{Gandy13, Lau13}), and a change pattern reporting rules engine.

The enhanced SPC-based algorithms use one or more detection thresholds to control the false detection rate. Detection thresholds are learned via simulation to account for sampling errors and possible serial dependence of change rates. For example, statistical time series models (e.g., ARMA, Gaussian white noise, etc.) for normalized rates of change could be used to simulate time series data for different KPIs under the hypothesis of no change. We run the SPC-based change detection algorithms over each simulated time series for each value in the set of trail thresholds. The fraction of detected cases (false alarm rate) are then computed for each trial threshold value. The threshold whose false alarm rate is closest to the target false alarm rate is identified.

\begin{figure}[h]	
	\centering	
	\includegraphics[width=\linewidth]{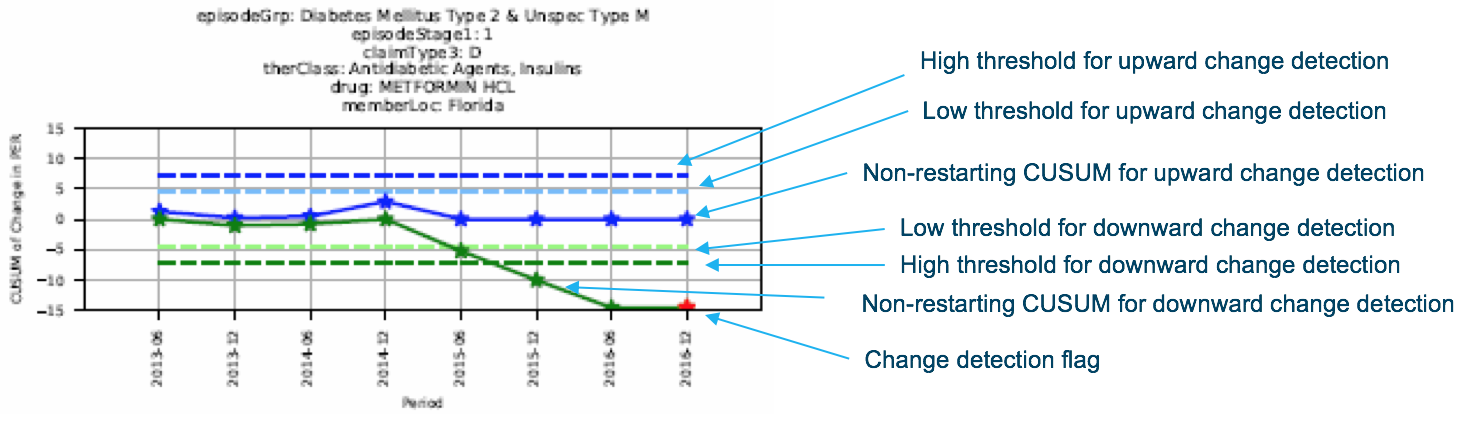}
	\caption{Example of Enhanced SPC Algorithms: Non-restarting CUSUM}
	\label{fig:cusum}
\end{figure}

The online detection algorithms automatically account for multiple change points. By specifying reporting rules, changes detected can be reported at any point within the analysis window, or reported at the end of the analysis to focus on the latest cumulative effect of change. Figure \ref{fig:cusum} shows an example of enhanced SPC algorithms with learned detection thresholds. High and low thresholds for upward and downward change detection are learned via simulation. In this example, the focus is on the latest cumulative effect of change. The non-restarting CUSUMs is applied here and the change detection flag is only triggered at the end of the analysis time period.

\begin{figure}[h]	
	\centering	
	\includegraphics[width=\linewidth]{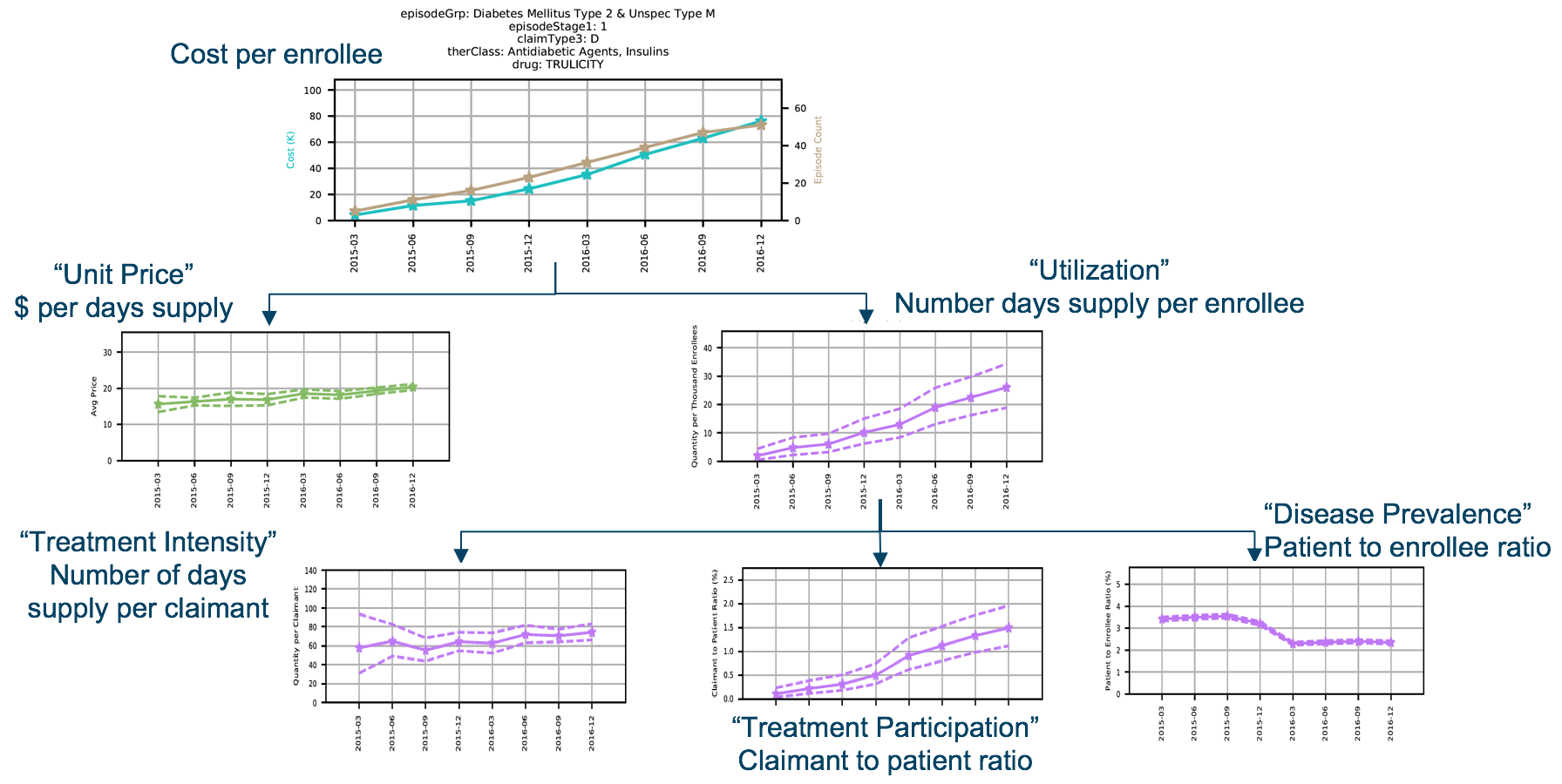}	
	\caption{Example of Multi-KPI Change Attribution}
	\label{fig:decomposition}
\end{figure}

Moreover, we attribute the impact of detected cost changes into multiple explanatory factors as shown in Figure \ref{fig:decomposition}.

The impact of change in cost per enrollee is determined by
$$c(t)=s(t)-s(t-T), t=T, \ldots, P$$
$$I(c)=EWA(c(T+1), \ldots, c(P))$$
where $s(t)$ is the cost per enrollee in period $t$, and EWA is the exponential weighted average $\frac{1-w}{1-w^{P-T}}\sum^{P}_{t=T+1}w^{P-t}c(t)$.

The total impact of change $I(c)$ could be decomposed by unit price $I(c_1)$ and utilization $I(c_2)$ as follows.
\begin{align*}
c_1(t)&=e(t-T)[a(t)-a(t-T)], t=T+1, \ldots, P\\
J(c_1)&=EWA(c_1(T+1), \ldots, c_1(P))\\
c_2(t)&=[e(t)-e(t-T)]a(t-T), t=T+1, \ldots, P\\
J(c_2)&=EWA(c_2(T+1), \ldots, c_2(P))
\end{align*}
where $e(t)$ is the quantity to enrollee ratio in period $t$, $a(t)$ is the unit price in period $t$.

Let $\delta_1=[J(c_1)+J(c_2)]-I(c)$, the impact due to unit price is
$$I(c_1)=J(c_1)-\delta_1\times|J(c_1)|/[|J(c_1)|+|J(c_2)|],$$
and the impact of change due to utilization is
$$I(c_2)=J(c_2)-\delta_1\times|J(c_2)|/[|J(c_1)|+|J(c_2)|].$$

Similarly, the impact of change due to utilization $I(c_2)$ could be further decomposed by treatment intensity, treatment participation and disease prevalence.

As an example, Figure \ref{fig:decomposition} shows the trends and confidence intervals of KPIs for Trulicity, a drug indicated for patients with type 2 diabetes. The KPI trends indicate that the cost increase of Trulicity is driven by an increase in utilization, particularly treatment participation. The total cost impact of Trulicity for treating diabetes has increased \$0.1087 per member per month (PMPM), among which only \$0.0098 (9.0\%) is due to unit price increase and the remaining \$0.0989 (91.0\%) is attributable to increases in utilization. Further decomposing the cost impact of utilization, treatment participation attributed to \$0.1044 increase PMPM, while treatment intensity only accounts for \$0.0073 increase. Conversely, the cost impact is reduced by \$0.0129 due to lower prevalence of types 2 diabetes over time.

\subsection{Characterize Change Patterns}\label{sec:characterize}
In order to turn the time-series results into interpretable themes or categorizations, we use a multi-resolution temporal aggregation scheme.

The change detection algorithm described in Section \ref{sec:detect} is executed on different analysis time windows (e.g., 2-year short-term vs. 5-year long-term window). Depending on the time windows, time series data are aggregated by different resolutions (e.g., 3-month vs. 6-month granularity for short and long run, respectively). Machine learning classifiers (e.g., rules-based, neural net, etc.\cite{Hastie09}) or unsupervised classifiers (e.g., clustering, etc.\cite{Hastie09}) could then be employed to assign shapes of change into characterized detection patterns.

Table \ref{tab:profile} shows an example using rules-based methods to characterize the change patterns according to the results of multi-resolution analysis. Short-run high resolution analysis and long-run low resolution analysis results are analyzed for various changes. As shown in Table \ref{tab:profile}, short-run and long-run results may reveal different change patterns, for example, emerging growth (recent increasing change) or persistent decline (long-term continued decreasing change).

\begin{table}	
	\caption{Rule-based Profiling for Change Pattern Characterization}	
	\label{tab:profile}	
	\begin{tabular}{ccl}		
		\toprule		
		Change patterns&Short time window&Long time window\\		
		\midrule		
		Emerging growth & increase & no change\\		
		Emerging decline & decrease & no change\\		
		Persistent growth & increase & increase\\		
		Persistent decline & decrease & decrease\\		
		Stabilizing growth & no change & increase\\		
		Stabilizing decline & no change & decrease\\		
		\bottomrule		
	\end{tabular}	
\end{table}

\section{Utilization Offsets}\label{sec:MOD}
%With change patterns detected in Section \ref{sec:detect}, we can also identify ``utilization offsets'' of comparable treatment options and quantify the cost impact attributable to the offset effects.
\subsection{Identify Offsetting Treatments}

The offsets are identified when utilization patterns of the comparable factors under the same condition move in opposite directions within the same time window.

Offsets may include drug treatments, place of services, disease severity, etc. Potential substitutions of drugs within the same therapeutic class and indication per published evidence may be analyzed using Micromedex RED BOOK database. Shifts in care settings may also be identified as well as shifts in disease severity stages.

\begin{figure}[h]	
	\centering	
	\includegraphics[width=\linewidth]{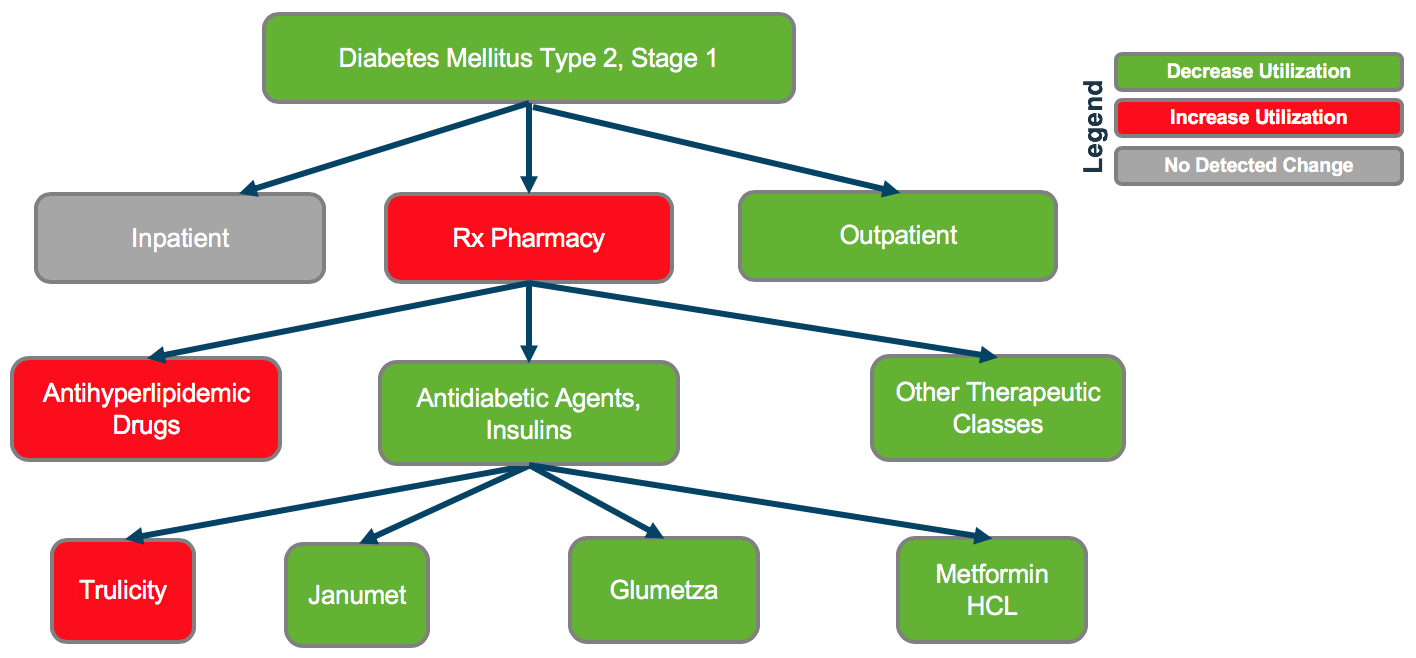}	
	\caption{Diabetes Type 2 Hierarchical Drill Paths}	
	\label{fig:offset}
\end{figure}

%For each medical condition, comparable treatment offsets with utilization trends moving in the opposite direction (MOD) are identified based on the hierarchical viewpoint. 
Figure \ref{fig:offset} shows an example of MOD identification for type 2 diabetes. The prevalence (patients to enrollee ratio) of diabetes type 2 is decreasing. An MOD group is identified with an increase in utilization for Rx Pharmacy, and a decrease in utilization for Outpatient services. This indicates a potential treatment shift from outpatient to drugs. Although the therapeutic class ``Antidiabetic Agents, Insulins'' shows a decrease in utilization, further drill downs reveal that the utilization of Trulicity is increasing while utilization of Janumet, Glumetza and Metformin HCL is decreasing. Note that Janumet and Glumetza are brand name drugs for Metformin HCL, which is also detected. Trulicity's active ingredient is a different compound (glucagon-like peptide-1 receptor agonist) and is administered via injection compared to the other oral medications. Our MOD analysis indicates a treatment shift for type 2 diabetes in this population and we quantify the impact of this observed shift.

\subsection{Estimate Treatment Offset Cost Impact}
Offset tracking can be accomplished at different granularities, from population-level to individual-level. Our approach focuses on estimating the impact using population-level data which reduces the analytic complexity.

We define an offsetting network including originators (treatments with decreased utilization) and receivers (treatments with increased utilization). %The offsetting network is not a closed-looped system. 
The utilization migration within the offsetting network and external factors are difficult to track at the population level; we cannot distinguish individuals who are existing versus new or discontinued patients for a treatment. We only have information on the total usage of the treatment at sub-population level, which poses challenges to accurately estimate the utilization migration. To determine the volume of utilization migration excluding the effects of new/discontinued patients, we postulate the following assumptions.

{\itshape Proportional Allocation Assumptions.} (a) The volume of outflow offset from a treatment option that experiences utilization decrease (originator) to a comparable treatment option that experiences utilization increase (receiver) is assumed to be proportional to the amount of observed utilization increase of the receiver. (b) The total volume of outflow offset from an originator is assumed to be proportional to the amount of observed utilization decrease of the originator.

{\itshape Migration Equilibrium Equations.} The offset inflow of a receiver is equal to the total offset outflow received from all of its originators.

{\itshape Maximum Migration Principle.} The outflow offset volumes from originators to receivers are determined by maximizing the total amount of outflow offset from all originators under the constraints of proportional allocation. The inflow offset volume of each receiver is determined by summing the outflow volumes from all originators.

%Denote $O_1, \ldots, O_I$ to be the originators, $R_1, \ldots, R_J$ to be the receivers of an offsetting network and $R(O_i)$ are all the receivers of originator $O_i$. 
Utilization migration for each treatment options within the network can be calculated based on Algorithm \ref{algo:MODimpact}. The cost impact due to offset can be computed by using offset-adjusted utilization under the assumption of no change in average cost. In our example, Trulicity substituting Metformin-based drugs resulted in a cost increase of \$0.2720 PMPM for treating type 2 diabetes.

\begin{algorithm}
	
	\DontPrintSemicolon % Some LaTeX compilers require you to use \dontprintsemicolon    instead
	
	\KwIn{An offseting network with outflow $o_1,\ldots, o_I$ from the originators $O_1, \ldots, O_I$ and inflow $r_1, \ldots, r_J$ to receivers $R_1, \ldots, R_J$ .}
	
	\KwOut{Offset outflow $o_{m,1},\ldots, o_{m,I}$ and offset inflow $r_{m,1},\ldots, r_{m,I}$.}
	
	Estimate the population of migrations that satisfies Maximum Migration Principle and the migration constrains	
	$P_m = \min\left[\sum_{i=1}^{I}o_i, 1/\left(\sum_{i=1}^{I}\frac{o_i}{\sum R(o_i)\sum_{i=1}^{I}o_i}\right) \right] $\\ %where $R(O_i)$ are all the receivers of originator $O_i$.\\
	
	Given migration population $P_m$, calculate within-network migration from each originators $o_{m, 1}, \ldots, o_{m, I}$ based on Proportional Allocation assumptions (b).\\
	
	Calculate migration transitions $(O_i \rightarrow R(O_i))$ based on Proportional Allocation assumptions (a).\\
	
	Calculate migration to each receivers $r_{m,1},\ldots, r_{m,I}$ based on Migration Equilibrium Equations.
	
	\caption{Algorithm of Calculating MOD Impact}
	
	\label{algo:MODimpact}
	
\end{algorithm}

\section{Conclusions}\label{sec:conclusion}

In this paper, we present a systematic approach that detects cost change drivers within large-scale healthcare claims data using hierarchical and multi-resolution search strategies. %Changes are detected based on year-to-year comparison of de-seasonalized and self-censored KPI change rates. An enhanced SPC algorithm is employed to automatically account for multiple change points (e.g., non-restarting CUSUM). Detection thresholds are learned via simulation or statistical theory to account for sampling errors and possible serial dependence of change rates. Change patterns are characterized by unsupervised clustering or rule-based techniques for better interpretability. Furthermore, the detected change patterns could be used to identify ``utilization offsets'' of comparable treatment options and to quantify their cost impacts.
 Our approach provides data scientists and payers an early-warning alert system to surface emerging cost drivers to support early intervention for healthcare cost management. We highlight changes in cost outcomes attributing these changes into actionable and clinically meaningful factors. This approach could also be applied to different settings including monitoring changes in healthcare quality or other performance-based metrics.

\bibliographystyle{ACM-Reference-Format}

\bibliography{WCD-KDD}

\end{document}